\documentclass[11pt]{article}
\newcommand{\Comment}[1]{{}}
\usepackage{amsfonts,amsthm,amsmath,amssymb,slashed}
\usepackage[textwidth = 430 pt, textheight = 630 pt]{geometry}
\usepackage{color} 
\usepackage{epsfig,latexsym}

\Comment{\usepackage{color}
\definecolor{MyDarkBlue}{rgb}{0.15,0.15,0.45}
\usepackage[linktocpage=true]{hyperref}
\hypersetup{
colorlinks=true,
citecolor=MyDarkBlue,
linkcolor=MyDarkBlue,
urlcolor=MyDarkBlue,
pdfauthor={Horatiu Nastase},
pdftitle={The title},
pdfsubject={hep-th}
}

\usepackage[numbers,sort&compress]{natbib}
\usepackage{hypernat}}
\usepackage{graphicx}
\usepackage{cite}

\newcommand\ignore[1]{}
\def\one{{\,\hbox{1\kern-.8mm l}}}

\def\Tr{{\rm Tr\, }}

\def\a{\alpha}\def\b{\beta}

\def\d{\partial}
\def\dag{\dagger}

\def\Tr{\mathop{\rm Tr}\nolimits}

\newcommand{\Cset}{{\,\,{{{^{_{\pmb{\mid}}}}\kern-.45em{\mathrm C}}}}}

\newcommand{\be}{\begin{equation}}
\newcommand{\bea}{\begin{eqnarray}}

\newcommand{\ee}{\end{equation}}
\newcommand{\eea}{\end{eqnarray}}

\begin{document}

\renewcommand{\thefootnote}{\fnsymbol{footnote}}

\makeatletter
\@addtoreset{equation}{section}
\makeatother
\renewcommand{\theequation}{\thesection.\arabic{equation}}

\rightline{}
\rightline{}




\begin{center}
{\LARGE \bf{\sc Towards deriving the AdS/CFT correspondence}}
\end{center} 
 \vspace{1truecm}
\thispagestyle{empty} \centerline{
{\large \bf {\sc Horatiu Nastase${}^{a}$}}\footnote{E-mail address: \Comment{\href{mailto:horatiu.nastase@unesp.br}}{\tt horatiu.nastase@unesp.br}}
                                                        }

\vspace{.5cm}


\centerline{{\it ${}^a$Instituto de F\'{i}sica Te\'{o}rica, UNESP-Universidade Estadual Paulista}} 
\centerline{{\it R. Dr. Bento T. Ferraz 271, Bl. II, Sao Paulo 01140-070, SP, Brazil}}

\vspace{1truecm}

\thispagestyle{empty}

\centerline{\sc Abstract}

\vspace{.4truecm}

\begin{center}
\begin{minipage}[c]{380pt}
{\noindent I present the sketch of a ``physicist's'' derivation of the AdS/CFT correspondence for the original pair, 
string theory in $AdS_5\times S^5$ vs. ${\cal N}=4$ SYM, based on the pp wave limit, and deviations from it. 
I show that one can reverse the logic, and derive the action of ${\cal N}=4$ SYM, as kinetic terms plus vertices, 
as well as the elements of its path integral, from 
string theory on the pp wave. One can also treat consistently deviations from the ``strings on the pp wave'' limit, 
starting from the SYM side, and a priori reconstruct the full $AdS_5\times S^5$ geometry perturbatively, as well as 
full perturbative string theory on it.

}
\end{minipage}
\end{center}

\vspace{.5cm}

\setcounter{page}{0}
\setcounter{tocdepth}{2}

\newpage

\renewcommand{\thefootnote}{\arabic{footnote}}
\setcounter{footnote}{0}

\linespread{1.1}
\parskip 4pt



\section{Introduction}

The AdS/CFT correspondence, introduced by Maldacena in \cite{Maldacena:1997re} 
(see the books \cite{Nastase:2015wjb,Ammon:2015wua} for more information), describes, in its original form, 
a relation between the string theory in an $AdS_5\times S^5$ background and ${\cal N}=4$ Super-Yang-Mills (SYM)
theory. 
The relation is obtained through taking a decoupling limit of a system of $N$ D3-branes, viewed in two possible ways:
as a gravitational solution in the low-energy supergravity limit of string theory, and as a string theory object on which 
open strings can end. This leads to a heuristic derivation of the proposed (conjectured) relation, in which the two theories
are found to be related, though the exact relation was at this time unknown, and was made concrete in 
\cite{Gubser:1998bc,Witten:1998qj}. The relation is holographic, and the SYM field theory is situated at the boundary 
of $AdS_5\times S^5$ in its Penrose diagram.
One also necessarily finds the relation in a certain `t Hooft limit, of large $N$ and
fixed and large $g^2_{YM}N$, corresponding to classical supergravity on the string theory side, but later the conjectured relation 
was found to hold at all $N$ and $g_{YM}$, through many explicit examples. 
After countless checks of the correspondence, the conjecture is now viewed as a virtual certainty, though a formal proof 
is still lacking, and the heuristic one is strictly speaking valid only for the classical supergravity limit of string theory. 

The first relation involving true string theory in the AdS/CFT correspondence was the relation between string theory 
on the pp wave obtained as a Penrose limit of $AdS_5\times S^5$, and the large-R-charge limit (sector of operators of 
large R-charge) of ${\cal N}=4$ SYM \cite{Berenstein:2002jq}, thereafter known as the BMN limit. The relation was 
later developed into the ``pp wave correspondence''. In this case the 
holographic nature of the relation was obscured, though it was clarified in \cite{Berenstein:2002sa}, but one had an 
explicit map between a discretized string theory worldsheet and a ``chain'' of large R-charge operators in a certain 
``dilute-gas approximation,'' later extended to a full spin chain in \cite{Minahan:2002ve}.

In this paper I show that one can take the pp wave correspondence and extend it to a sketch of a derivation of the AdS/CFT 
correspondence. First, I will show that we can reverse the logic of  \cite{Berenstein:2002jq}, and not only derive the 
discretized string action on the pp wave from the SYM, but also derive the action of SYM, as well as its 
perturbation theory, from the string action on the 
pp wave. Second, I will show that, based on the analysis of corrections in  \cite{Berenstein:2002sa}, we can extend the 
results away from the pp wave limit and the free string action. The new ingredient is the proof that corrections away 
from the pp wave background can be found from corrections away from the large $g^2_{YM}N$ limit in SYM. 

Of course, I will not show explicitly that the partition functions of the two sides of the relations agree, only that the
ingredients for the perturbation theory on both sides can be found from the other one. But since the relation is a duality, 
and one perturbation theory corresponds to a non-perturbative statement on the other, the probability that in this way 
we miss some important non-perturbative ingredient is vanishingly small, and we have the (sketch of) a ``physicist's 
derivation'' of the AdS/CFT correspondence. 

The paper is organized as follows. In section 2, I will derive the SYM action (or Hamiltonian) from the free string on the pp wave. 
In section 3, I will show that the gauge group is uniquely determined, and we obtain the correct path integral. In section 4, I will 
analyze the corrections on the string theory side: corrections on the pp wave in $1/\mu^2$, string (field theory) corrections in $g_s$, and 
$\a'$ corrections in the free string action, away from the pp wave and into AdS, and how they can appear from SYM. In section 5, I will focus
on the last, $\a'$ corrections to the free string action, away from the pp wave and into AdS, and translate the result in SYM language. In section 6, I will derive 
these corrections from the SYM diagrams, and in section 7 I conclude. In Appendix A, I will review how the Cuntz Hamiltonian construction, needed for the $1/\mu^2$
corrections, appears from the SYM, and in Appendix B, I will review how the string (field theory) corrections in $g_s$ appear from SYM. 

\section{From string action on the pp wave to SYM Hamiltonian}

In  \cite{Berenstein:2002jq} it was shown that a SYM calculation for the anomalous dimension (or rather, $\Delta-J$)
of (later so-called) ''BMN 
operators'' in the large $J$ limit reproduces the lightcone energy $p^-$ of the string on the pp wave obtained as a Penrose 
limit of $AdS_5\times S^5$. Thus the free string on the pp wave is obtained from SYM theory, which is the ``$\Rightarrow$''
part of the equivalence at the free level. We will not review it here, as it is well known. 

But another important element that was derived in  \cite{Berenstein:2002jq} was a discretized version of the string action on 
the pp wave, which was matched against a calculation of a ``Cuntz oscillator'' Hamiltonian obtained from the SYM large-$J$
operators, viewed as a spin chain. In this section we will expand on this construction, and show that it can be turned into 
a derivation of the SYM Hamiltonian from the string action, thus completing the ``$\Leftarrow$'' part of the equivalence 
at the free level. 

The pp wave obtained as Penrose limit of $AdS_5\times S^5$ is 
\be
ds^2=2dx^+dx^--\mu^2(\vec{r}^2+\vec{y}^2)(dx^+)^2+d\vec{y}^2+d\vec{r}^2\;,
\ee
where $\vec{r}$ is a 4-dimensional Euclidean space obtained by scaling the spacelike part of $AdS_5$, and $\vec{y}$ 
is a 4-dimensional Euclidean space obtained by scaling the part of $S^5$ transverse to the direction of motion of the 
null geodesic around which we expand.

We will analyze the matching of expansion parameters, and what it means, in detail in the next sections. For the moment, 
we will just  define as a {\em map} between the parameters of the string on the pp wave (and away from it), 
$\a'\mu p^+$, $g_s$ (and $\a'/R^2$), and the SYM parameters $g_{YM}, N$ and $J$,
\be
\a'\mu p^+=\frac{J}{\sqrt{g^2_{YM}N}}\;,\;\; 
4\pi g_s=g^2_{YM}\;,\;\;
\left(\frac{\a'}{R^2}=\frac{1}{\sqrt{g^2_{YM}N}}\right).
\ee

The string action on the pp wave, in lightcone gauge, was derived in \cite{Berenstein:2002jq}, and is 
simply a free action with a mass term for the 8 worldsheet scalars, plus a fermionic part required by supersymmetry. 
For simplicity, we will focus on the bosonic part, while the fermionic part can always be recovered by supersymmetry. 
It is
\be
S=\sum_{I=1}^8\int d\tau \int_0^L d\sigma \frac{1}{2}[(\dot \phi^I)^2-(\phi'^I)^2-(\phi^I)^2]\;,
\ee
where the scalars $\phi^I$ come from the spacetime coordinates $\vec{r}$ and $\vec{y}$ and the string length is
\be
L=2\pi \a' p^+=\frac{2\pi J}{\mu \sqrt{g^2_{YM}N}}\propto \frac{1}{\left[\frac{g^2_{YM}N}{J^2}\right]^{1/2}}.
\ee

The large $J$ charge limit corresponding to the Penrose limit keeps the parameter $\frac{g^2_{YM}N}{J^2}$ fixed, so 
the string length in units of $\mu$ is fixed. 

The string Hamiltonian on the pp wave is thus 
\be
H=\sum_{I=1}^8\int_0^L d\sigma \frac{1}{2}[(\dot \phi^I)^2+(\phi'^I)^2+(\phi^I)^2].
\ee

We want to discretize this string Hamiltonian in order to obtain a SYM counterpart. The length $L$ of the string will be 
divided into $J$ units, where 
\be
J=p^+R^2\;,
\ee
and $R$ is the radius of $AdS_5$ and $S^5$, from which the pp wave is derived (we have to remember that 
we will include next AdS corrections to the pp wave, so the AdS parameters exist in the theory). 
The discretization is then defined as 
\be
\phi^I(x,t)\rightarrow \phi_l^I\equiv \mu \tilde \phi_l^I\equiv \mu \frac{a_l^Ie^{i\omega t}+a_l
^{\dagger I} e^{-i\omega t}}{\sqrt{2}}\;,
\ee
where we put $\omega=1$.

With this recipe, we obtain 
\be
(\phi^I)^2+(\dot \phi^I)^2=\left( \frac{a_l^Ie^{i\omega t}+a_l^{\dagger I} e^{-i\omega t}}{\sqrt{2}}\right)^2
-\omega^2\left( \frac{a_l^Ie^{i\omega t}-a_l^{\dagger I} e^{-i\omega t}}{\sqrt{2}}\right)^2
=a_l^I a_l^{\dagger I} +a_l^{\dagger I} a_l^I\;,
\ee
and for the derivative (we have defined, as usual, the 't Hooft coupling $\lambda=g^2_{YM}N$)
\be
\phi'^I\rightarrow \frac{\phi_{l+1}^I-\phi_l^I}{L/J}=\frac{\mu \sqrt{\lambda}}{2\pi}(\tilde \phi_{l+1}^I-\tilde \phi_l^I).
\ee

Finally then, the discretized string Hamiltonian on the pp wave is 
\be
\frac{H}{\mu^2}=\sum_{I=1}^8\sum_{l=1}^J\left[\frac{a_l^{\dagger I} a_l^I+a_l^I a_l^{\dagger I}}{2}
+\frac{\lambda}{2(2\pi)^2}(\tilde \phi_{l+1}^I-\tilde \phi_l^I)^2\right].\label{stringHam}
\ee  

The interacting part of the above Hamiltonian will describe the interacting part of the SYM Hamiltonian for the scalars. 
The dimension is a problem at first sight, as the discretized Hamiltonian is in 0+1 dimensions, but the SYM theory is in 
3+1 dimensions, or more precisely in 4 Euclidean dimensions, since we consider the Wick rotation of the space. 
However, using the conformal invariance of ${\cal N}=4 $ SYM, we map the Euclidean 
$\mathbb{R}^4$ space on which it lives to $\mathbb{R}_t\times S^3$, and then we KK reduce the theory on the $S^3$. 
It is this reduced theory that will be obtained from the discretized string Hamiltonian on the pp wave. 

It is known that the symmetries of string theory on $AdS_5\times S^5$ match against those of ${\cal N}=4 $ SYM. 
We will not need to discuss the full supergroup symmetries. For our purposes, it suffices to note that the $SO(4,2)$ isometry
group of $AdS_5$ becomes the conformal group of ${\cal N}=4 $ SYM, the $SO(6)$ isometry of $S^5$ the R-symmetry
group of ${\cal N}=4$ SYM, and the 8 4-dimensional supersymmetries of string theory in $AdS_5\times S^5$ become
the 4 supersymmetries plus 4 conformal supersymmetries of ${\cal N}=4$ SYM. This justifies the use of conformal invariance
to turn $\mathbb{R}^4$ into $\mathbb{R}_t\times S^3$ and KK reduce to $S^3$. It also means that, strictly speaking, 
we only need to derive the scalar propagator and interaction in the SYM Hamiltonian, the other terms being obtained 
from them by ${\cal N}=4 $ supersymmetry. 

Further, we consider the symmetries of the full $AdS_5\times S^5$ when analyzing the above pp wave Hamiltonian, 
and split the 8 massive fields $\phi^I$ of the string action into 4 $\phi^m$, $m=1,...,4$ fields  that came from the 
$S^5$ directions, and 4 $D_a$, $a=1,...,4$ fields that came from the $AdS_5$ directions. In the string action on the 
pp wave, the difference is not obvious, but in the AdS corrections away from the pp wave, that will be considered later, 
it is, so we can assume that the split is done. 

Consider now the time-independent interaction piece in the discretized string Hamiltonian (\ref{stringHam}),
\be
\frac{H_{\rm int, t-indep.}}{\mu^2}=\frac{\lambda}{4(\pi)^2}\sum_{I=1}^8\sum_{l=1}^J[2(a_j^Ia_j^{\dagger I}
+a_j^{\dagger I} a_j^I)-(a_j^I a_{j+1}^{\dagger I}+a_{j+1}^{\dagger I} a_j^I +a_{j+1}^Ia_j^{\dagger I} 
+a_j^{\dagger I} a_{j+1}^I)].\label{intHam}
\ee

Now instead of operators $a_j^I$ acting on (the Hilbert space at) a site $j$ on a dilute gas approximation, close to the 
global vacuum, $|0\rangle_1\otimes |0\rangle_2\otimes...\otimes |0\rangle_k\otimes...$, consider a redefinition of the states
and Hamiltonian, such that we have a string  of {\em Cuntz oscillators} $b^\dagger$ and $a^{\dagger I}$, 
with $|0\rangle_k$ standing in for an operator $b^\dagger$ inserted at position $k$ inside a ``word''. 
Cuntz oscillators $a_i$, for $i=1,...,n$, are oscillators that satisfy the algebra
\be
a_i|0\rangle=0\;, \;\;\;
a_i a_j^\dagger=\delta_{ij}\;,\;\;\;
\sum_{i=1}^n a_i^\dagger a_i=1-|0\rangle\langle 0|.
\ee

Since there are no relations between $a_i^\dagger a_j^\dagger$ and $a_j^\dagger a_i^\dagger$, the order matters, and a 
string of Cuntz oscillators corresponds to a certain ``word'' made up of the different $a_i^\dagger $'s. 
But as noted in \cite{Gopakumar:1994iq}, this is just the Hilbert space of $n$ large $N$ random matrices acting on a 
vacuum. Note that in reality, for the transformation from $a_j^I$ to $(a^I,b)_j$'s to be precise, we need the $a_j^I$'s to be derived also from a 
sort of Cuntz oscillators at a site, but when acting on state in the dilute gas approximation, the difference between the two 
vanishes as $J\rightarrow \infty$, as found in  \cite{Berenstein:2002jq,Cardona:2014ora}.

The analysis of the Cuntz oscillator Hamiltonian was implicit in \cite{Berenstein:2002jq}
and described explicitly in Appendix A of \cite{Cardona:2014ora}. For completeness, I review it here also in Appendix A.

When translating the action of the interaction Hamiltonian (\ref{intHam}) onto the same redefined Hilbert space, 
we see that the effect of $a_{j+1}^{\dagger I} a_j^I$ is to shift the position of an $a^I$ oscillator in a sea of $b$ oscillators
to the right ($j\rightarrow j+1$), whereas the effect of $a_j^{\dagger I} a_{j+1}^I$ is the same shift to the left
($j+1\rightarrow j$), and the other terms subtract the case when nothing happens, leaving a commutator interaction. 
Note that the terms with time dependence in (\ref{stringHam}) are needed to turn (\ref{intHam}) into a relativistic field formulation, but otherwise we can 
work with (\ref{intHam}). 

Finally splitting $\phi^I$ into $\phi^m$ and $D_a$ as promised, we arrive at the interacting Hamiltonian
\be
H_{\rm int}=2\lambda\left(\sum_{m=1}^4 [b^\dagger, \phi^m][b,\phi^m]
+\sum_{a=1}^4 [b^\dagger, D_a][b,D_a]\right).\label{Hint}
\ee

Here we should remember that the map to Cuntz oscillators meant that we actually have large $N$ matrices. We can 
isolate an $N$ out of $\lambda$, and call the rest $g^2_{YM}$ (more on that in the next section). Moreover, we 
understand $b^\dagger $ as creating a complex scalar $Z$ (also a large $N$ matrix) in the dilute gas approximation. 
We also ``undo'' the KK reduction (``oxidize'') on $S^3$ and think of $D_a$ as a covariant derivative in the 
$\mathbb{R}^4$ that is conformal to $\mathbb{R}_t\times S^3$. Finally then, the interaction Hamiltonian in 4 dimensions
is
\be
H_{\rm int}=-g^2_{YM}\Tr\left[\sum_{m=1}^4[Z,\phi^m][\bar Z,\phi^m]+\sum_{a=1}^4[D_a,Z][D_a,Z^\dagger]\right]\;,
\ee
and it is understood that it refers to large $N$ matrix fields. 

So we have a large $N$ matrix theory with the full interacting Hamiltonian obtained from the above by the acting
with ${\cal N}=4$ supersymmetry. Stricty speaking, $D_a$ refers to a covariant derivative, so it contains also the kinetic
term for the scalars, which means that the ${\cal N}=4$ supersymmetry will derive all the kinetic terms as well. 

However, we will be a bit redundant, and argue for the scalar kinetic term also in another way. 

The SYM interacting Hamiltonian above acts on a discrete ``string'' (or ``spin chain'') of operators in the dilute gas
approximation, with $J$ scalars $Z$ or, in other words, with R-charge $J$. This is so, since we have the  map 
$p^+\leftrightarrow U(1)_J$, relating the isometries of $AdS_5\times S^5$ in the Penrose limit with the symmetries 
(conformal plus R) of ${\cal N}=4$ SYM in the large charge limit  \cite{Berenstein:2002jq}. The other element of the map 
is $H=p^-\leftrightarrow \Delta - J$, where $\Delta$ is conformal dimension, and it acts on the operators. 

Since SYM is a conformal field theory, operators of definite $\Delta $ and $J$, and moreover with 
$\Delta - J\ll J$ (from the Penrose limit/large R-charge limit)  must be gauge invariant, and defined at some 
point $x$. Then the {\em free} result of the correlator of two gauge invariant operators has a factor $1/|x-y|^{2J}$, 
which means that one field has the propagator $\propto 1/|x-y|^2$. We write the large $N$ planar diagrams for these
operators using double lines, for $(i\bar j)$ pairs (having $q$ and $\bar q$ at the sides of the thick lines), which means that 
in addition to the $1/|x-y|^2$ factor, we must have also delta functions for the double indices, in other words the 
propagator is
\be
P_{i\bar j}^{k\bar l}(x,y)=\frac{\delta_i^k\delta_{\bar j}^{\bar l}}{(x-y)^2}.
\ee

This amounts to having the canonical kinetic term for the scalars, in the adjoint representation of $SU(N)$.

\section{Gauge group and path integral}

Next, we need to understand the gauge group, which we claimed to be $SU(N)$. First, up to now we have only 
$\lambda$, defined from the discretized string theory as $[J/(\mu \a' p^+)]^2$. In the $J\rightarrow \infty$ limit, 
we then consider $g^2_{YM}=4\pi g_s$ in terms of the string coupling, and the remainder is $N\rightarrow \infty$, 
which at least approximates an integer (since it is close to infinity). Then the standard choice of gauge group is $SU(N)$. 
Needing a complex representation rejects $SO(N)$, but moreover we know that $SO(N)$ and $USp(2N)$ appear 
in string theory only from orbifold and orientifold projections, which is not the case here. This takes care of all the classical 
groups of arbitrarily large rank, so we have uniquely defined the gauge group. 

Thus we have derived the SYM action from the free string theory on the pp wave. But to complete the picture of the 
SYM theory, we need to understand that the path integral is also correct. However, as is well known from quantum field 
theory, in the Feynman diagrammatic expansion (via the Wick and Feynman theorems), having the correct measure for the 
path integral translates into proper integration and summation over Feynman diagrams. For instance, a 
famous example making that 
point more poignantly is the fact that the Fujikawa method for the path integral anomaly becomes the one-loop anomaly 
in Feynman diagrams, properly integrated over. 

But that was implicit in the one-loop Feynman diagram calculation in  \cite{Berenstein:2002jq} that reproduced the 
first correction of the string energy on the pp wave. While here I have shown how to derive the interacting Hamiltonian 
from the string one, one can reversely calculate, perturbatively, in Feynman diagrams, the corrections to the string energy 
resulting from the SYM Hamiltonian. While doing so, we need to consider a 
Feynman diagram with a single interaction between two operators at points $x$ and $y$. It gives a contribution of 
\be
\sim \int d^dz \frac{1}{(z-x)^4(z-y)^4}\;,
\ee
compared with the $1/(y-x)^4$ of the free result. This is however only in the case of the correct integration over the 
Feynman diagram (and the correct summation of diagrams). 

The same logic continues to apply when we calculate corrections to the free string on the pp wave, as we will do in the 
next sections. For instance, one can consider the correction due to a SYM vertex to the free string splitting diagram
 (the vertex being at the split), 
where ${\cal O}(x)$ splits into ${\cal O}(y)$ and ${\cal O}(z)$, and which gives a contribution of 
\be
\sim \int d^dw\frac{1}{(w-x)^4(w-y)^2(w-z)^2}\;,
\ee
compared to the free result of $1/[(x-y)^2(x-z)^2]$. 

On top of this argument, we consider the fact that the `t Hooft planar limit at large $N$ is correlated with the existence of the 
Haar measure over the $SU(N)$ gauge group.

Thus the correct path integration is also implicit in the map between the string on the pp wave and SYM. But we 
need to go beyond the free string on the pp wave, which is why we turn to corrections next.

\section{Matching expansion parameters, and corrections}

Let us recapitulate what we have learned so far, and see what remains to be done. We have obtained the free string on the pp wave from SYM (in the original 
BMN paper), and the SYM action and path integral from strings on pp waves (here). Since the correspondence is between SYM and strings on $AdS_5\times S^5$, 
we need to show also that corrections away from the pp wave, constructing $AdS_5\times S^5$ perturbatively, are derived from the SYM. That will be done in the 
next two sections. But we also need to show that the correspondence works at arbitrary $N$ and $g_{YM}$ (on the SYM side), or $\a'$ and $g_s$ (on the string side). 
Since we derived the SYM {\em action, } including the vertex, and path integral, the correct SYM theory at arbitrary $N$ and $g_{YM}$ exists on one side, so 
we need the elements of the correct string theory at arbitrary $\a'$ and $g_s$ on the other: the string ($g_s$) vertex on the pp wave and pp wave corrections in $\a'$
(and corrections away from the pp wave).  

The dimensionless expansion parameters on both sides are:

\begin{itemize}

\item on the string side: $g_s$ and $1/(\mu\a' p^+)$ on the pp wave, and $\a'/R^2$ in AdS (away from the pp wave). 

\item on the SYM side: $g_{YM}$ and $N$ defining the theory, and $J$ defining the subset of operators (the BMN sector). 

\end{itemize}

Matching the two sides, we have the correspondence of expansion parameters as follows:

\begin{itemize}

\item There is a dimensionless parameter on the pp wave, matched to the ("bare", or "as obtained directly from the pp wave") coefficient of the SYM vertex {\em in the BMN limit},
\be
\frac{1}{(\mu \a' p^+)^2}=\frac{g^2_{YM}N}{J^2}.
\ee
This was obtained from SYM in \cite{Berenstein:2002jq}.

\item The parameter defining string field theory corrections on the pp wave, which was obtained from SYM in \cite{Berenstein:2002sa},
\be
(4\pi g_s)^2(\a'\mu p^+)^4=\frac{J^4}{N^2}=\frac{1}{(N/J^2)^2}=\left(\frac{g_{YM}^2}{\frac{g_{YM}^2N}{J^2}}\right)^2.
\ee
In SYM, this parameter defines {\em nonplanar corrections}. This was described in \cite{Berenstein:2002sa} and is reviewed in Appendix B.

\item Finally, there is the parameter describing deviations from the pp wave, and into $AdS_5\times S^5$, 
\be
\frac{\a'}{R^2}=\frac{1}{\sqrt{g^2_{YM}N}}.
\ee
It describes corrections away from the strong coupling limit, more precisely away from the 't Hooft limit $g^2_{YM}N=$ fixed and large, on the SYM side. 
This parameter was not addressed yet, and will be addressed in the next two sections. 

\end{itemize}

Some more observations on the first two expansion parameters. The first one appeared in the string action we wrote, since
\be
\frac{g^2_{YM}N}{J^2}=\left(\frac{2\pi}{\mu L}\right)^2\;,
\ee
and thus helped us define the SYM vertex. 

The second parameter defines string (field theory) corrections, since the splitting of a discretized string into two (discrete "pair of pants" diagram), 
corresponding to an order $g_s$ diagram, gives an order 
\be
\frac{J^2}{N}=(4\pi g_s)(\a' \mu p^+)^2
\ee
in SYM. 

In conclusion, we can build perturbatively the string expansion on the pp wave, and if the last parameter works, also on the $AdS_5\times S^5$.

\section{String action corrections away from the pp wave}

To show that the last parameter works as well perturbatively, I will show that string action corrections away from the pp wave can be obtained from SYM. 
In this section I calculate the string action corrections away from the pp wave, and write it in terms of SYM quantities. 

We first write the $AdS_5\times S^5$ metric in global coordinates to the first nontrivial order (beyond the pp wave),
\bea
ds^2&=&R^2[-\cosh^2 \rho d\tau^2+d\rho^2+\sinh^2\rho d\Omega_3^2]+R^2[\cos^2\theta d\psi^2+d\theta^2
+\sinh^2\theta d\Omega_3'^2]\cr
&\simeq & R^2\left[-\left(1+\rho^2+\frac{\rho^4}{3}\right)d\tau^2+d\rho^2+\left(\rho^2+\frac{\rho^4}{3}\right)d\Omega
_3^2\right]\cr
&&+R^2\left[\left(1-\theta^2+\frac{\theta^4}{3}\right)d\psi^2+d\theta^2+\left(\theta^2-\frac{\theta^4}{3}\right)
d\Omega_3'^2\right].
\eea
With the usual rescalings of 
\be
\tilde x^\pm =\frac{\tau\pm \psi}{\sqrt{2}}\;,\;\; 
\tilde x^+=x^+\;,\;\;
\tilde x^-=\frac{x^-}{R^2}\;,\;\; 
\rho=\frac{x}{R}\;,\;\;
\theta=\frac{y}{R}\/,
\ee
we obtain the metric
\bea
ds^2&\simeq & -2dx^+dx^-\left(1+\frac{x^2-y^2}{2R^2}\right)-(dx^+)^2\left[\frac{x^2+y^2}{2}+\frac{x^4-y^4}{2R^2}
\right]\cr
&& +dx^2+x^2d\Omega_3^2+\frac{x^4}{3R^2}d\Omega_3^2+dy^2+y^2d\Omega_3'^2-\frac{y^4}{3R^2}
d\Omega_3'^2.
\eea
With the further rescalings $x^+\rightarrow \sqrt{2}\mu x^+$, $x^-\rightarrow x^-/(\sqrt{2}\mu)$, we obtain 
finally 
\bea
ds^2&\simeq & -2dx^+dx^-\left(1+\frac{x^2-y^2}{2R^2}\right)-\mu^2(dx^+)^2\left[x^2+y^2
+\frac{x^4-y^4}{R^2}
\right]\cr
&& +d\vec{x}^2+\frac{x^4}{3R^2}d\Omega_3^2+d\vec{y}^2-\frac{y^4}{3R^2}
d\Omega_3'^2.
\eea

Now we can write the (Polyakov) string action in this background, which is the action on the pp wave plus corrections,
\bea
S&=&-\frac{1}{2\pi \a'}\int d\tau\int_0^L d\sigma\frac{1}{2}\sqrt{-\gamma}\gamma^{ab}\left[-2\d_a X^+\d_b X^-
\left(1+\frac{x^2-y^2}{2R^2}\right)\right.\cr
&&\left.-\mu^2\d_a X^+ \d_b X^+\left(\vec{x}^2+\vec{y}^2+\frac{x^4-y^4}{2R^4}\right)
\right.\cr
&&\left.+\d_a x^i \d_b x^i +\d_a y^i\d_b y^i +\frac{x^4}{3R^2}\d_a \Omega_3\d_b \Omega_3
-\frac{y^4}{3R^2}\d_a \Omega_3'\d_b \Omega_3'\right].
\eea

In the lightcone gauge 
\be
\sqrt{-\gamma}\gamma^{ab}=\eta^{ab}\;,\;\;
X^+=\tau\;,
\ee
we obtain 
\bea
S&=&-\frac{1}{2\pi \a'}\int d\tau \int_0^{2\pi \a' p^+}d\sigma \left[\frac{1}{2}\d_a x^i \d^a x^i+\frac{1}{2}\d_a y^i\d^a 
y^i+\frac{\mu^2}{2}\left(x^2+y^2+\frac{x^4-y^4}{R^2}\right)\right.\cr
&&\left.+\frac{x^4}{3R^2}(\d_a\Omega_3)^2-\frac{y^4}{3R^2}(\d_a \Omega_3')^2\right].
\eea

Defining dimensionless variables $\tilde x\equiv x/\sqrt{\a'}$ and $\tilde y\equiv y/\sqrt{\a'}$, and writing the 
deviation from the pp wave action, we get
\be
\delta S\equiv S-S_{\rm pp} =-\left(\frac{\a'}{R^2}\right)\frac{1}{2\pi}\int d\tau \int_0^{2\pi \a'p^+}d\sigma \left[\frac{\mu^2}{2}
(\tilde x^4-\tilde y^4)+\frac{\tilde x^4}{3}(\d_a \Omega_3)^2-\frac{\tilde y^4}{3}(\d_a \Omega_3')^2\right].\label{deltaS}
\ee
Here the $\tilde y$ directions correspond to the scalars $\phi^m$ and the $\tilde x$ directions to the spacetime covariant derivatives
$D_a$, and as advertized the split of the pp wave coordinates into $\tilde y$ and $\tilde x$  is not arbitrary anymore when considering the first corrections away from the 
pp wave. 

Next we write this result in terms of SYM quantities, and write it as corrections away from the 't Hooft limit, in $1/\sqrt{g^2_{YM}N}$. 

Since the action (\ref{deltaS}) is complicated, we consider it first together with the $\mu\rightarrow \infty $ limit, or more precisely, 
the $\mu \a' p^+=1/\sqrt{\frac{g^2_{YM}N}{J^2}}\rightarrow \infty $ limit, which can be thought of also as the $g^2_{YM}N$ fixed, 
$J\rightarrow \infty$ limit.  That means that we are still in the case of infinitely long operators, but now for finite $g^2_{YM}N$, which 
means that there are now other states of finite (yet very large) mass that should be considered. In the strict BMN limit, these states had infinite mass, and 
decoupled from the calculation.

In this further limit, the extra term in the action (\ref{deltaS}) becomes 
\be
\delta S=-\left(\frac{\a'}{R^2}\right)\frac{1}{4\pi}\int d(\mu \tau) \int_0^{2\pi \mu \a' p^+}d(\mu \sigma)(\tilde x^4-
\tilde y^4).
\ee

We see that indeed, this gives a 
\be
\frac{\a'}{R^2}=\frac{1}{\sqrt{g^2_{YM}N}}
\ee
correction in SYM.

Next, translating this $\delta S$ into a variation of the energy or Hamiltonian, noting that  for the static case, $\delta S_{\rm string}=-\delta H\int d\tau$, 
we see that the SYM change in anomalous dimension, $\delta \Delta=\delta(\Delta -J)\leftrightarrow \delta H$ 
(and ignoring the $\mu$ factors that translate from SYM to string dimensions) is
\bea
\delta\Delta&=&+\frac{1}{\sqrt{\lambda}}\frac{1}{4\pi}\int_0^{2\pi \mu \a' p^+}d(\mu \sigma)(\tilde x^4-\tilde y^4)
=\frac{1}{\sqrt{\lambda}}\frac{1}{4\pi}2\pi \mu \a' p^+(\tilde x^4-\tilde y^4)\cr
&\rightarrow &\frac{1}{4\pi \sqrt{\lambda}}\frac{2\pi J}{4\sqrt{\lambda}}\sum_l [(a_l^a+a_l^{\dagger a})^4-(a_l^m+a_l^{\dagger m})
^4]\cr
&=&\frac{J}{8\lambda}\sum_l [(a_l^a+a_l^{\dagger a})^4-(a_l^m+a_l^{\dagger m})^4].\label{deltaDelta}
\eea
Here we have used the same map from the pp wave, $(\tilde x, \tilde y)\rightarrow \left(\frac{(a_l^a+a_l^{\dagger a})}
{\sqrt{2}},\frac{(a_l^m+a_l^{\dagger m})}{\sqrt{2}}\right)$, just that now, as 
we see, there is no equivalence between $\tilde x$ and $\tilde y$ anymore. Of course, the relation (\ref{deltaDelta}) is formal only, we started with a relation where 
$\tilde x,\tilde y$ were numbers, and $\delta \Delta $ also, and ended up with a relation for operators, in effect going from the eigenvalue difference $\delta\Delta$
to a Hamiltonian operator $\hat H$.

\section{SYM calculation matching string action corrections}

In this section, I show that we can obtain the correction to the anomalous dimension (\ref{deltaDelta}) from the contribution of very massive states, with masses
of order $g^2_{YM}N$. The existence of these operators (states) with large anomalous dimension (mass) was already discussed in \cite{Berenstein:2002jq}.
BMN operators have insertions of $\phi^m$ or $D_a$ inside the "vacuum" $\Tr [Z^J]$, and these fields have $\Delta-J=1$. But for the field $\bar Z$, we have 
$\Delta-J=2$, so an operator with it as an insertion a priori could mix with an operator with two insertions of $\phi^m$ or $D_a$. 

Consider then the two kinds of operators, that can a priori mix, 
\bea
{\cal O}_1^{m,n}&=&\frac{1}{\sqrt{J}}\sum_{l_1,l_2}\frac{1}{\sqrt{J}N^{\frac{J+2}{2}}}
\Tr[Z_0 Z^{l_1}\phi^m Z^{l_2}\phi^n Z^{J-l_1-l_2}]
e^{\frac{2\pi i l_1 n_1}{J}}e^{\frac{2\pi i n_2 (l_1+l_2)}{J}}\cr
{\cal O}_2&=&\frac{1}{\sqrt{J}}\sum_l \frac{1}{N^{\frac{J+2}{2}}}\Tr[Z_0 Z^l \bar Z Z^{J-l}]e^{\frac{2\pi i n l}{J}}\;,
\eea
where $Z_0$ is some arbitrary insertion (different than $Z$), put in the first position so that it fixes the cyclicity of the trace. 
We will consider both the case $m=n$ and the case $m\neq n$ in the following.

Then, as noted in  \cite{Berenstein:2002jq}, the ``decay amplitude'' for $\bar Z\rightarrow \phi \phi$ (the classical anomalous 
dimension is the same between the two) comes from the one-loop planar 
contribution to the $\langle \bar{\cal O}_1^{mm}(x){\cal O}_2(y)\rangle
$ correlator, generated by the $Z\bar Z\phi^m\phi^m$ vertex,
and which is of order $\lambda=g^2_{YM}N$ (note that for this process we have necessarily $m=n$, otherwise the one-loop diagram
vanishes). But since $\lambda$ is infinite, there is really no perturbation theory 
for this object, and all loops contribute, so we can't really say the amplitude is of order $\lambda$, only that it blows up when $\lambda$ blows up. 

Similarly, for the ``energy'' (anomalous dimension minus charge) of the $\bar Z$ excitation, we have the one-loop planar 
contribution to the correlator $\langle {\cal O}_2(x){\cal O}_2(y)\rangle$ defined by the same vertex $Z\bar Z \phi^m\phi^m$, which again is of order $\lambda$, and again 
we cannot conclude that the energy is of this order, since we don't have perturbation theory, and all the loops contribute. 

In this section we want to show that the mixing of operators ${\cal O}_1$ and ${\cal O}_2$, which was negligible in the strict $\lambda\rightarrow \infty$ limit, 
is nonzero at finite $\lambda$, and contributes to the anomalous dimension $ \Delta_1 $ of ${\cal O}_1$, with a contribution matching the one calculated for the 
string in the AdS-corrected pp wave in the previous section. 

I first review the formalism describing the mixing of operators (in the BMN limit), that is reasonably well known. 
For a general 2-point function of operators (in the BMN large $J$ charge limit), 
\bea
\langle {\cal O}_\a (x) \bar{\cal O}_\b (0)\rangle &\simeq&\frac{S_{\a\b}}{|x|^{2(J+2)}}+\frac{T_{\a\b}}{|x|^{2(J+2)
+2\Delta_A}}\cr
&=&\frac{1}{|x|^{2(J+2)}}(S_{\a\b}+T_{\a\b}\log|x\Lambda|^{-2})\;,
\eea
where we write spacetime-independent operators ${\cal O}_\a$ and define
\be
S_{\a\b}=\langle {\cal O}_\a \bar{\cal O}_\b\rangle\;, \;\;\;
T_{\a\b}=\langle {\cal O}_\a H \bar {\cal O}_\b\rangle.
\ee

The dilatation operator $D$, calculating the anomalous dimension, is defined by 
\be
D\cdot {\cal O}_\a={D_\a}^\b {\cal O}_\b\;, \;\;\;
{D_\a}^\b=(J+2)\delta_\a^\b +T_{\a\gamma}(S^{-1})^{\gamma\b}.\label{dilat}
\ee

We want to diagonalize $D$, so that we can calculate a well-defined anomalous dimension for an eigenstate (as the diagonal element of $D$). We call the 
diagonal operators (with the mixing diagonalized) ${\cal O}'_A$. Then we have
\be
D\cdot {\cal O}'_A=\Delta_A{\cal O}'_A\Rightarrow D\cdot {\cal O}_A=(V_{\a A}\Delta_A V^{-1}_{A\b}){\cal O}_\b\;,\label{dilatati}
\ee
such that we obtain the correlators
\be
\langle {\cal O}_\a(x)\bar {\cal O}_\b(0)\rangle=V_{\a A}V^*_{\b B}\langle {\cal O}'_A(x) {\cal O}'_B(0)\rangle
=V_{\a A}V^*_{\b B}\frac{\delta_{AB}C_A}{|x|^{2(J+2)+2\Delta_A}}\;,
\ee
where in the second equality we have assumed that ${\cal O}'_A$ are diagonal, i.e., proportional to $\delta_{AB}$, and of well-defined anomalous dimension
$\Delta_A$.

Equating (\ref{dilat}) with (\ref{dilatati}), we obtain the equations
\bea
{D_1}^1&=& (V_{11})^2\Delta_1+(V_{12})^2\Delta_2=(J+2)+H_{11}\cr
{D_1^2}={D_2}^1&=&V_{11}V_{12}\Delta_1+V_{12}V_{22}\Delta_2=H_{12}\cr
{D_2}^2&=&(V_{22})^2\Delta_2+(V_{21})^2\Delta_1=(J+2)+H_{22}\;,
\eea
where
\be
H_{\a\b}=\langle {\cal O}_\a H {\cal O}_\b\rangle=T_{\a\b}.
\ee

Note that {\em at one loop} we need ${\cal O}_1^{mm}$ ($m=n$) in order to have a nonzero $H_{12}=T_{12}$, but 
at higher loops we could have $m\neq n$. Since we are interested in the all-loop result, we should keep general at this 
point.

Normalizing as $V_{11}=V_{22}=1$, we obtain the equations
\bea
&&\Delta_1+(V_{12})^2\Delta_2=J+2+H_{11}\cr
&&\Delta_2+(V_{12})^2\Delta_1=J+2+H_{22}\cr
&&V_{12}(\Delta_1+\Delta_2)=H_{12}.
\eea

Eliminating $V_{12}$ and replacing back in the remaining equations, we get
\bea
&& \Delta_1+\Delta_2\left(\frac{H_{12}}{\Delta_1+\Delta_2}\right)^2=J+2+H_{11}\cr
&& \Delta_2+\Delta_1\left(\frac{H_{12}}{\Delta_1+\Delta_2}\right)^2=J+2+H_{22}.
\eea

Until now, the formalism was general, but now we specialize to the limiting case of interest for us.

If $H_{22}\gg H_{11}$, so that $H_{12}/(\Delta_1+\Delta_2)\ll 1$, 
\bea
&&  \Delta_1+\left(\frac{H_{12}}{\Delta_1+\Delta_2}\right)^2(J+2+H_{22})\simeq J+2+H_{11}\Rightarrow\cr
&&\Delta_1\simeq J+2+H_{11}-\frac{[(J+2+H_{22})H_{12}^2}{2(J+2)+H_{11}+H_{22}]^2}.
\eea

If moreover $H_{22}\gg J$, then we obtain for the anomalous dimension of operator ${\cal O}_1$, corrected by the mixing with operator ${\cal O}_2$,
\be
\Delta_1\simeq J+2+H_{11}-\frac{H_{12}^2}{H_{22}}.
\ee

This result is the analog of the see-saw mechanism, by which a matrix $\begin{pmatrix} 0 & M\\ M & B\end{pmatrix}$, 
with eigenvalues
\be
\lambda_\pm =\frac{B\pm \sqrt{B^2-4M^2}}{2}
\ee
has, if $B\gg M$, the eigenvalues $B$ and $-\frac{M^2}{B}$, with the latter being the smallest. One could call our result then a 
sort of "see-saw for anomalous dimension".

Next we remember that actually there are two kinds of insertions in ${\cal O}_1$: with $\phi^m$ and with $D_a$. To understand them, we write the relevant 
vertices in ${\cal N}=4$ SYM. With the conventions
\bea
F_{ab}&=&\d_aA_b-\d_b A_a+g[A_a,A_b]\cr
\Tr[T_AT_B]&=&-\frac{1}{2}\delta_{AB}\cr
[T_A,T_B]&=&{f_{AB}}^CT_C\cr
D_a&=&\d_a+g[A_a,.]\cr
A_a&=&A_a^AT_A\cr
\phi_m&=&\phi_m^AT_A\;,
\eea
the SYM action has the terms
\bea
S_{SYM}&=& (-2)\Tr\int d^4x \left\{-\frac{1}{4}F_{ab}^2-\frac{1}{2}D_aZ\overline{D_aZ}-\frac{g^2}{4}([\phi_m,\phi_n]^2
+2[\phi_m,Z][\phi_m, \bar Z])\right\}+...\cr
&=&(-2)\Tr\int d^4x \left(-\frac{g^2}{4}\right)\left\{[A_a,A_b]^2+[\phi_m,\phi_n]^2+2[A_a,Z][A_a,\bar Z]\right.\cr
&&\left.+2[\phi_m,Z][\phi_m,\bar Z]\right\}+...\cr
&&
\eea
that give the 4-point interactions needed in the diagrams:

\begin{itemize}

\item the $[Z,\phi_m][\bar Z,\phi_m]$ vertex gives both the BMN vertex for $\phi_m$ insertion, and the decay of $Z,\bar Z
\rightarrow \phi_m \phi_m$. 

\item the vertex $[A_a,Z][A_a,\bar Z]$ gives both the BMN vertex for $D_a Z$ insertion, and the decay of $Z,\bar Z\rightarrow
D_a Z \overline{D_a Z}$.

\end{itemize}

That means that the one-loop vertex $H_{11}$ with insertion $D_a Z$ is the {\em same } as $H_{11}$ for $\phi^m$ insertion, 
and moreover, the mixing $H_{12}$ of ${\cal O}_2$ is, not just with ${\cal O}_1^{mn}$, but also with the same operator ${\cal O}_1$, but with 2 $D_a Z$ insertions
(since the first order $H_{12}$ transition, or {\em decay} is the same for both insertions). However, there is a difference: 
the two $A_a$'s have to be adjacent to form the one-loop vertex for decay of $Z\bar Z$, and that only occurs for 
$-(D_aZ)(D_aZ)=-[A_a,Z][A_a,Z]=[Z,A_a][A_a,Z]$, so we have a minus sign difference between the two operators.

Next, we consider what happens with the mixing when we have mixing of two types of ${\cal O}_1$ with ${\cal O}_2$. 
Then, for two ${\cal O}_1$ operators, 1 for $\phi^m$ and 1' for $D_a$, mixing with ${\cal O}_2$, we find:
\bea
\Delta_1&=&J+2+H_{11}-\Delta_2 \frac{H_{12}^2}{(\Delta_1+\Delta_2)^2}\simeq 
J+2+H_{11}-\frac{H_{12}^2}{\Delta_2}\cr
\Delta_{1'}&=&J+2+H_{1'1'}-\Delta_2 \frac{H_{1'2}^2}{(\Delta_{1'}+\Delta_2)^2}\simeq 
J+2+H_{1'1'}-\frac{H_{1'2}^2}{\Delta_2}\cr
\Delta_2&=&J+2+H_{22}-\Delta_1\frac{H_{12}^2}{(\Delta_1+\Delta_2)^2}-\Delta_{1'}\frac{H_{1'2}^2}{(\Delta_1+
\Delta_2)^2}\simeq H_{22}.\label{mixingtwo}
\eea

Until now the see-saw type correction to the anomalous dimension we have obtained doesn't look much like the string result. Moreover, as we said, $H_{12}$ and $H_{22}$
are actually nonperturbative in the large $g^2_{YM}N$ limit, unlike $H_{11}$, which is perturbative in $g^2_{YM}N/J^2$. However, I will argue that the 
$H_{22}/H_{12}^2$ ratio is not only perturbative, but calculable through a simple factorization argument, and the result matches the string calculation. More precisely, 
I want to show that 
\be
H_{22}=H_{12}^2\frac{8\lambda}{J}\;,
\ee
which gives the right contribution to $\delta \Delta$ for the $\phi$ insertions via the see-saw mechanism.

The idea is that $H_{22}$ is a transition between two ${\cal O}_2$ operators, which can be factorized over an insertion of a
${\cal O}_1^{mn}$ operator in the middle. Note that generically the one-loop contribution to 
$H_{22}$ will be subleading with respect to higher loop ones, as we said, but the factorization is considered {\em for the full, all-loop amplitude}. On the other hand, 
the propagation of the ${\cal O}_1$ operator is {\em perturbative}, as we were just reminded, so {\bf is} given by the one-loop 
result, which is in this case, of order
\be
\frac{g^2_{YM}N}{J^2}\times J=\frac{\lambda}{J}.
\ee

Indeed, the leading contribution to $H_{11}$ is from a single vertex acting on only one of the $\phi^m$ insertions in ${\cal O}_1$
(we would naively say that the leading contribution is the free part, without any vertices, but we remember that $H$ is defined as the operator with the free part 
subtracted). Then we have $\lambda/J^2$ from the vertex, and a factor of $J$ from the normalization. To be more precise, the factorization is 
\be
\langle {\cal O}_2(x) H {\cal O}_2(0)\rangle \simeq \sum_j \langle {\cal O}_2(x){\cal O}_1^{mn}(y)\rangle_j\langle  
{\cal O}_1^{mn} (y) H {\cal O}^{mn}_1(y)\rangle_j \langle {\cal O}_1^{mn}(y) {\cal O}_2(0)\rangle\;,
\ee
with 
\be
 \langle {\cal O}_1^{mn}(y) {\cal O}_2(0)\rangle\sim 
  \langle {\cal O}_1^{mn}(y) H {\cal O}_2(0)\rangle.
\ee

An analysis of the factors of $J$ in sums and normalizations is as follows. ${\cal O}_2$, appearing at the two extremes 
of the factorization, has a $\bar Z$ in a position summed over, giving a factor of $J$, cancelling the $(1/\sqrt{J})^2$ from the 
normalization of the two operators ${\cal O}_2$. There is a single sum in this case, since the $Z_0$ in the two ${\cal O}_2$'s
is aligned to be on the first position in both (in the free contraction, the two would be contracted). Then for the 
${\cal O}_1  H {\cal O}_1$ contraction, the two $\phi$'s need to be in the same position, and the two $Z_0'$'s of the two 
${\cal O}_1$s also in the same position. Note that one of the $\phi$'s is contracted free, and the other one 
is contracted with the $H$ vertex. 
That means that there are 2 sums over $J$, for the positions of the two $\phi$'s, cancelling the $(1/J)^2$ from the 
normalization of the two ${\cal O}_1$ operators. 

But there is still the sum over $J$ since in the factorization, the position of the $Z_0$ (the origin) of the ${\cal O}_2$
operator need not match the position of the $Z_0$ (origin) of the ${\cal O}_1$ operator, as the $H_{12}$ 
element is {\em nonperturbative} (so not bound by the approximately free contraction, like the middle $H_{11}$ one is). 
Thus we get  the sum over $J$ leading to the factor of $J$. 

Next, we consider the factorization in the presence of mixing with two operators ${\cal O}_1$ and ${\cal O}_{1'}$. 
The mixing was defined in (\ref{mixingtwo}), and the factorization becomes 
\be
\Delta_2\simeq H_{22}\simeq \sum_i H_{i2}^2 H_{ii}\;,
\ee
where in principle $i$ is everything, but only the lowest energy states contribute in the leading approximation, i.e., the 
states with one excitation, which have nonzero $H_{i2}$ (the vacuum has $H_{i2}=0$). Then we obtain 
\bea
\delta \Delta_1&=&-\frac{H_{12}^2}{\sum_i H_{i2}^2 H_{ii}}\cr
\delta \Delta_{1'}&=&-\frac{H_{1'2}^2}{\sum_i H_{i2}^2 H_{ii}}.
\eea

This is understood as an insertion of the identity written as $\sum_n |n\rangle\langle n|=1$, which reduces inside the VEV to the sum over $|{\cal O}_1\rangle
\langle \bar{\cal O}_1|$, as the one of lowest energy. Note that really we have 
an insertion of $|{\cal O}_1\rangle\langle \bar {\cal O}_1| |{\cal O}_1\rangle\langle\bar{\cal O}_1$ so, because of the 
complex conjugation, even a general phase inserted in ${\cal O}_1$, much less a minus sign as we saw there is in ${\cal O}_{1'}$, 
doesn't change the matrix element. Then we must have
\be
H_{1'1'}=H_{11}\;,\;\; H_{1'2}^2=H_{12}^2.
\ee

Yet, from the string result, we saw that we need a relative minus sign between the terms for ${\cal O}_1$ and for ${\cal O}_{1'}$. This is crucial, since 
the two will build up the $AdS_5$ and the $S^5$ parts of the metric, with the different signs building up the $\cosh\rho$ vs. $\cos r$ factors. 

However, the minus sign comes when considering the correspondence with the string action, and the needed Wick rotation, as follows.

First, note that the operators in the formal relation (\ref{deltaDelta}) can be replaced by fields in SYM through the correspondence
(from the leading, pp wave, calculation)
\be
\frac{1}{4}\sum_l \left[(a_l^i+a_l^{\dagger i})^4-(a_l^m+a_l^{\dagger m})^4\right]
\rightarrow \left[\sum_a D_iZ(x) D_i \bar Z(x)\right]^2-\left(\sum_m \phi_m(x)\phi_m(x)\right)^2.
\ee
That means that really, only in the last step, of replacing the Hamiltonian matrix element with a Hamiltonian itself, do we need to obtain a relative minus sign between 
the $D_i Z$ contribution and the $\phi^m$ contribution. More precisely, this last step involves putting the creation operators (that give the matrix elements) 
next to the numerical contribution of (the sum is over 8 oscillators)
\be
\delta \Delta=-\frac{H_{12}^2}{\sum_i H_{i2}^2 H_{ii}}=-\frac{J}{8\lambda}.
\ee

But, as mentioned, $D_aZ$ are Euclidean, whereas the pp wave itself is Minkowskian, so we must have a Wick rotation. The Wick rotation needs to be the one common 
for the boundary of AdS space (AdS space becomes the pp wave in the Penrose limit) in global coordinates, i.e., the Wick rotation of $\mathbb{R}_t\times S^3$. In terms 
of the flat space conformal to it, this is a {\em radial time Wick rotation}, corresponding to the replacement
\be
D_r Z\rightarrow i D_r Z\;,
\ee
where $r$ is the radial direction of Euclidean space. 

Then for the KK reduction on $S^3$ implicit in the quantum mechanical Hamiltonian notation, the radial Wick rotation corresponds to the replacement
\be
D_a Z D_a \bar Z\rightarrow -D_aZ D_a \bar Z\;,
\ee
generating the required minus sign. 

That means that indeed, we have obtained the corresponding correction to the pp wave Hamiltonian implicit in the string formula (\ref{deltaDelta}). 

This was the last step of the derivation, that showed that also the perturbations of the string action away from the pp wave and into AdS space match with the 
SYM result. 

To recapitulate, I have shown that:

\begin{itemize}

\item the SYM action and perturbative path integral are derived from the free string theory in the pp wave. That could in principle leave some doubt about the 
nonperturbative contributions to SYM, so we needed to understand the corrections corresponding to it, away from infinite coupling, from the string theory side. 

\item all the 3 types of corrections on the string theory side: string (field theory) corrections in $g_s$, pp wave strong field corrections in $1/\mu^2$ and $\a'$
corrections away from the pp wave and into AdS, are reproduced from SYM. 

\end{itemize}

This completes the sketch of a "physicist's derivation" 
of the basic correspondence between ${\cal N}=4 $ SYM and string theory on $AdS_5\times S^5$.

\section{Conclusions}

In this paper I have shown a (not fully rigorous, so a "physicist's") derivation 
of the basic AdS/CFT correspondence. I have not shown explicitly the equality of the 
two partition functions, $Z_{\rm string, AdS}$ and $Z_{\rm CFT}$, which would have been the fully rigorous proof, but which would necessitate a fully nonperturbative 
definition of string theory, which is not available. Given our current tools, the most one could hope for was what I did: to derive the SYM action and path integral 
components from string theory, and to show that all possible perturbative corrections in string theory are reproduced in SYM. Of course, I have only shown how to 
deal with the basic (leading) string correction, not with the full perturbative expansion, which again would necessitate an understanding of the latter, which is 
not currently available. In other words, I have gone as far as current technology in string theory allows one to go towards a fully rigorous proof of the correspondence.

In this paper, I have considered only the original case of $AdS_5\times S^5$ vs. ${\cal N}=4$ SYM, and the maximal 
supersymmetry was needed, to obtain the full ${\cal N}=4$ SYM action from the scalar vertex. In other cases, with less
supersymmetry, like for instance the ABJM \cite{Aharony:2008ug} case, the argument of this paper cannot be straightforwardly used, hence 
we didn't address it.

\section*{Acknowledgements}
I would like to thank Juan Maldacena for discussions.
My work is supported in part by CNPq grant 304006/2016-5 and FAPESP grant 2014/18634-9. I would also 
like to thank the ICTP-SAIFR for their support through FAPESP grant 2016/01343-7, and the University of Cape Town for hospitality during the time that this project
was finished.

\appendix

\section{Cuntz oscillator Hamiltonian and its diagonalization}

In this appendix, I will review the calculation implicit in \cite{Berenstein:2002jq}
and described explicitly in Appendix A of \cite{Cardona:2014ora}.

The boundary of global AdS space is $\mathbb{R}_t\times S^3$, conformal to $\mathbb{R}^4$, on which SYM lives. To reach the quantum mechanical Hamiltonian, 
one KK dimensionally reduces on $S^3$ the SYM fields, and keeps the constant modes, while the higher KK modes correspond to the fields with insertions of 
the covariant derivatives $D_a$ (in the $\mathbb{R}^4$ picture). But because of the conformal curvature coupling to the scalars, even $Z$ by itself has a frequency 
of one, and has a corresponding oscillator of ${(b^\dagger)^\a}_\b$ (here $\a,\b$ are $SU(N)$ indices). On the other hand, $\phi^m, D_a Z$ have a frequency of two, 
and a corresponding oscillator of ${(a^\dagger)^\a}_\b$ (where we have suppressed the $m,a$ indices on $a^\dagger$), 

As explained in the text, the vacuum of the large $J$ charge sector of SYM is represented, in this KK reduction on $S^3$, as $\Tr[(b^\dagger)^J]|0\rangle\equiv |0\rangle_J$.
But in general, for an arbitrary set of oscillators $a^\dagger$ and $b^\dagger$ acting on $|0\rangle$, forming a {\em word} because of the random $SU(N)$ matrix 
structure, the Hilbert space was found, in \cite{Gopakumar:1994iq}, to correspond to the Hilbert space of Cuntz oscillators, with modified oscillator rules
\be
a_i|0\rangle=0\;\;\; a_ia_j^\dagger=\delta_{ij}\;\;\; \sum_{i=1}^n a_i^\dagger a_i=1-|0\rangle\langle0|\label{cuntzosc}
\ee
and no other relations (in particular no relations between $a^\dagger_ia^\dagger_j$
and $a^\dagger_ja^\dagger_i$, so that the order is important). The Cuntz oscillator formalism does not take into account the cyclicity of the trace, which must be imposed
by hand, otherwise the random large $N$ $SU(N)$ matrix structure is completely described by it.

Note then that for {\em a single Cuntz oscillator}, one has
\be
a|0\rangle=0;\;\;\; aa^\dagger=1\;\;\; a^\dagger a=1-|0\rangle\langle0|\;,
\ee
with the number operator
\be
\hat{N}=\frac{a^\dagger a}{1-a^\dagger a}=\sum_{k=1}^{\infty} (a^\dagger)^ka^k.
\ee

The next simplification is to consider the "dilute gas approximation", with a very large number $J$ of $b^\dagger $ oscillators corresponding to $Z$, and only a few $a^\dagger$
oscillators corresponding to $\phi^m$ and $D_aZ$. In this approximation, we can neglect the possibility of two $a^\dagger$ oscillators being next to each other (as being 
subleading to order $1/J$), and we can forget the $b^\dagger$ oscillators, and instead replace them with $J$ "sites" on a spin chain, and {\em at each site} consider 
an independent $a^\dagger_j$ Cuntz oscillator, where $j=1,..,J$ labels the sites. Each of the oscillators would obey the one-Cuntz oscillator algebra above, 
\be
a_ia_i^{\dagger} =1, \;\;\; a_i^{\dagger} a_i=1-(|0\rangle\langle 0|)_i;\;\; a_i|0\rangle_i=0\;,\label{cuntzbi}
\ee
and the oscillators are independent in that the ones for different sites commute
\be
[a_i, a_j]=[a_i^{\dagger}, a_j]=[a_i^{\dagger}, a_j^{\dagger}]=0 ,\;\;\; i\neq j.
\ee

Next, since on the pp wave string theory side, the oscillators correspond to momentum modes along the closed string, the reasonable thing to do is to the same on the 
SYM spin chain side, and consider the "Fourier" modes around the (cyclic) chain,
\be
a_j=\frac{1}{\sqrt{J}}\sum_{n=1}^Je^{\frac{2\pi i jn}{J}} a_n.
\ee

Doing so, for these momentum modes Cuntz oscillators, we obtain the commutation relations (modified algebra of oscillators)
\be
[a_n, a_m^{\dag}]=\frac{1}{J}\sum_{j=1}^Je^{\frac{2\pi ij(m-n)}{J}}
(|0\rangle\langle 0|)_j;\;\;\; [a_n,a_m]=[a_n^\dagger,a_m^\dagger]=0.
\ee

This looks complicated, however, since we are on the "dilute gas approximation", with states
\be
|\psi_{\{ n_i\} }\rangle=|0\rangle_1...|n_{i_1}\rangle...|n_{i_k}\rangle...|0\rangle_J\;,
\ee
we can write approximately 
\be
[a_n, a_m^{\dagger}]|\psi_{\{ n_i\} }\rangle=\left(\delta_{nm} -\frac{1}{J}
\sum_k e^{2\pi i i_k\frac{m-n}{J}} \right)|\psi_{ \{ n_i \} }\rangle\;,
\ee
where the second, $1/J$ term, is a correction in this approximation. Ignoring it, we obtain the usual commutation relations $[a_n,a_m^\dagger]=\delta_{mn}$.
We also obtain $a_n|0\rangle=0$, which means that $a_n,a_n^\dagger$ act as usual creation and annihilation operators in the dilute gas approximation.

As explained in the text also, the interaction Hamiltonian of SYM, understood as (\ref{Hint}), is turned into an action on each site $j$, by a field $\phi_j=(a_j+a_j^\dagger)/\sqrt{2}$.
One obtains part of it from the commutator interaction which is a result of the one-loop Feynman diagrams, and the rest are obtained from the need to construct the 
relativistic field $\phi_i$. Adding the free oscillator part (remember that $a_i^\dagger$ are oscillators with frequency of two), and reinstating the removed indices $I=m,a$,
we obtain ($\tilde \lambda =g^2_{\rm YM}N/(2\pi)^2$ here)
\be
\frac{H}{\mu^2}=\sum_{I=1}^8\sum_{j=1}^J\left[\frac{a_j^Ia_j^{\dagger I}+a_j^{\dagger I}a_J^I}{2}+\tilde \lambda\left(a_{j+1}^I+a_{j+1}^{\dagger |}-a_j^I-a_j^{\dagger I}\right)^2\right]
\ee

We observe that 
\be
H|0\rangle={\rm const.}|0\rangle+2\mu^2\tilde \lambda \sum_{I=1}^8\sum_{j=1}^J \left(a_{j+1}^{\dagger I} a_j^{\dagger I}-a_j^{\dagger I} a_j^{\dagger I}\right) |0\rangle\neq 0\;,
\ee
which contradicts the fact that $\Tr[Z^J]\rightarrow |0\rangle_J$ is a good BPS vacuum. But we assume that supersymmetry cures this problem through fermion loops, 
and redefine by hand $H$ to $\tilde H$ such that $\tilde H|0\rangle=0$, so we put $H|0\rangle =0$ by hand in the following.

We next write the Hamiltonian in terms of the Fourier modes $a_n^{\dagger I}$. We further make another transformation suggested by the string theory side
(but which can be argued for directly in the spin chain): because momentum must be conserved, the physical mode must be a mode with one excitation of momentum $+n$, 
and one of momentum $-n$. We thus redefine further
\bea
a^I_{n}&=&\frac{c_{n,1}^I+c^I_{n,2}}{\sqrt{2}}\nonumber\\
a^I_{J-n}&=&\frac{c^I_{n,1}-c^I_{n,2}}{\sqrt{2}}\;,
\eea
and substituting in the Hamiltonian, we obtain 
\bea
\frac{H}{\mu^2}&=&\sum_{I=1}^8\sum_{n=1}^{[J/2]}\left[\frac{c_{n,1}^{\dag I}c^I_{n,1}+c^I_{n,1}c_{n,1}^{\dagger I}}{
2}+\frac{c_{n,2}^{\dagger I}c^I_{n,2}+c_{n,2}^Ic_{n,2}^{\dagger I}}{2}+\right.
\nonumber\\
&&\left.\alpha_n(c^I_{n,1}+c_{n,1}^{\dagger I})^2-\alpha_n(c^I_{n,2}-c_{n,2}^{\dagger I})^2
+\beta_n[c^I_{n,1}-c_{n,1}^{\dagger I}, c^I_{n,2}+c_{n,2}^{\dagger I}] \right]\;,
\eea
where we have defined
\bea
\alpha_n&=&2\tilde \lambda \left(\cos \frac{2\pi n}{J}-1\right)= -4\tilde \lambda \sin ^2\frac{\pi n}{J}  
\nonumber\\
\beta_n&=& 2i \tilde \lambda \sin \frac{2\pi n}{J}.
\eea

However, considering the commutation relations for the momentum oscillators $a_n^{\dagger I},a_n^I$ in the dilute gas approximation, the commutator term in the above 
is found to vanish, and we obtain 
\bea
\frac{H}{\mu^2}&=&\sum_{I=1}^8\sum_{n=1}^{[J/2]}\left[\frac{c_{n,1}^{\dag I}c^I_{n,1}+c^I_{n,1}c_{n,1}^{\dagger I}}{
2}+\alpha_n(c^I_{n,1}+c_{n,1}^{\dagger I})^2\right.
\nonumber\\
&&\left. +\frac{c_{n,2}^{\dagger I}c^I_{n,2}+c_{n,2}^Ic_{n,2}^{\dagger I}}{2}
-\alpha_n(c^I_{n,2}-c_{n,2}^{\dagger I})^2\label{decoupled}
 \right]\;,
\eea
so the two modes $c_{n,1}^I$ and $c_{n,2}^I$ have decoupled. These two modes are perturbed oscillators, which can be diagonalized by a Bogoliubov 
transformation. 

Indeed, a  Hamiltonian of the form 
\be
H=\frac{aa^{\dagger}+a^{\dagger} a}{2}\pm \frac{\mu^2}{2}\frac{(a\pm a^{\dagger})^2}{2}
=\left(1+\frac{\mu^2}{2}\right)\frac{aa^{\dagger}+a^{\dagger}a}{2}\pm \frac{\mu^2}{4}
(a^2+{a^{\dagger}}^2)\;,\label{generic}
\ee
under the Bogoliubov transformation 
\bea
b&=&\tilde \alpha a \pm \tilde\beta a^{\dagger}\nonumber\\
\tilde \alpha -\tilde \beta &=& 1/\sqrt{\omega}\;\;\; \tilde \alpha +\tilde\beta =\sqrt{\omega}
\nonumber\\
\omega&=&\sqrt{1+\mu^2}
\eea
becomes
\be
H=\omega\frac{bb^{\dagger}+b^{\dagger}b}{2}.
\ee

Applying this to (\ref{decoupled}), we obtain 
\be
\frac{H}{\mu^2}=\sum_{I=1}^8\sum_{n=1}^{J/2}\omega_n\left[\frac{\tilde{c}_{n,1}^{\dagger I}\tilde{c}_{n,1}^I+
\tilde{c}_{n,1}^I\tilde{c}_{n,1}^{\dagger I}}{
2}+\frac{\tilde{c}_{n,2}^{\dagger I}\tilde{c}_{n,2}^I+\tilde{c}_{n,2}^I\tilde{
c}_{n,2}^{\dagger I}}{2}\right]\;,
\label{hami}
\ee
where the Bogoliubov transformation on the oscillators is
\bea
&& \tilde{c}_{n,1}^I=\tilde \a_nc_{n_1}^I+\tilde \b_nc_{n,1}^{\dagger I}\nonumber\\
&&\tilde{c}_{n,2}^I=\tilde \a_nc_{n_1}^I-\tilde \b_nc_{n,1}^{\dagger I}\nonumber\\
&&\tilde\a_n=\frac{(1+\alpha_n)^{1/4}+(1+\alpha_n)^{-1/4}}{2}\nonumber\\
&&\tilde\b_n=\frac{(1+\alpha_n)^{1/4}-(1+\alpha_n)^{-1/4}}{2}\;,
\eea
and the resulting energy of the eigenstates of the Hamiltonian is
\be
\omega_n=\sqrt{1+4\alpha_n}=\sqrt{1+4 \tilde\lambda \sin^2 \frac{\pi n}{J}}
=\sqrt{1+\frac{4g_sN}{\pi} \sin^2 \frac{\pi n}{J}}.
\ee

One observation is that the Bogoliubov transformation acts on the Cuntz oscillators and changes it, even though in the dilute has approximation $a_n^I$ had 
approximately normal commutation relations (a Bogoliubov transformation on normal oscillators maintains their algebra). In particular, for normal oscillators, 
$a|0\rangle=0$ implies
$b|0\rangle\neq 0$, and $b|0'\rangle=0$ gives $|0'\rangle=\exp(-\beta (a^\dagger)^2/\alpha)|0\rangle$. But the latter would violate the dilute gas approximation, 
so in fact now the vacuum has still zero energy. 

The second observation is that the calculation was exact in $\lambda$ and in $n/J$, in the $J\rightarrow \infty $ limit and the dilute gas approximation. 
If however we restrict to $n\sim 1\ll J$, we obtain the result in \cite{Berenstein:2002jq},
\be
\omega_n\simeq \sqrt{1+\tilde\lambda \left(\frac{2\pi n}{J}\right)^2}=\sqrt{1+g_sN \frac{4\pi n^2}{J^2}}.
\ee

The final observation is that the result only considered one-loop diagrams, i.e., order $\lambda$ ones, so the fact that we have an exact result in $\lambda$ 
implies that the Bogoliubov transformation, generating the square root, corresponds to a sort of resumming of successive one-loop contributions.

\section{String field theory corrections from SYM}

In this appendix, I will review the way to obtain string field theory (in $g_s$) corrections from SYM, described in \cite{Berenstein:2002sa}.
I will not review the full analysis of \cite{Berenstein:2002sa}, since it is very extensive, and detailed, but just a few salient points about the 
closed string interactions in the pp wave. In fact, since we have already shown that the SYM action and path integral is derived from the free string, I will only 
review the fact that there is a systematics of SYM diagrams that reproduces the systematics of the closed string expansion. 

First off, in SYM in the large $J$ charge limit we have, as we saw, the expansion parameters $\frac{g^2_{YM}N}{J^2}$, corresponding to $1/(\mu\a' p^+)^2$, 
so to large pp wave background, and $J^4/N^2=(4\pi g_s)^2(\mu\a' p^+)^4$, so to string field theory corrections, besides the 't Hooft expansion $1/\sqrt{g^2_{YM}N}$, 
corresponding to $\a'/R^2$ corrections away from the pp wave. That means that there are actually several large charge sectors, in terms of the length $J$ of SYM operators, 
compared with combinations of $g_{YM}$ and $N$. 

\begin{itemize}

\item The regime where $J\ll g_{YM}\sqrt{N}$ (which implies also $J\ll \sqrt{N}$), yet still parametrically comparable. In SYM, this corresponds to large $\frac{g^2_{YM}N}{J^2}$ 
corrections (so resummation is needed), but small $\frac{J^4}{N^2}$ corrections, so small nonplanarity. In the pp wave, this corresponds to small $\mu\a' p^+$ background, 
near flat space, and also small $g_s(\mu \a' p^+)^2$, so small string field theory corrections. At still lower $J$, we have the usual AdS/CFT regime.

\item The regime of $g_{YM}\sqrt{N}\ll J\ll \sqrt{N}$ is where we start our expansion in this paper. In SYM, this means small $\frac{g^2_{YM}N}{J^2}$ corrections, as well as small 
nonplanarity $\frac{J^4}{N^2}$. In the pp wave, this corresponds to large RR background, $\mu\a' p^+\gg 1$, but small string field theory corrections, so $g_s(\mu\a' p^+)^2\ll 1$. 

\item At still larger $J$, ($g_{YM}\sqrt{N}\ll$) $\sqrt{N}\ll J\ll \sqrt{N}/g_{YM}$, we have $\frac{g^2_{YM}N}{J^2}\ll 1$, so small perturbative SYM corrections, but $\frac{J^4}{N^2}
\ll 1$, so large nonplanarity (nonplanar diagrams dominate). In the pp wave, we have large RR background, $\mu\a' p^+\gg 1$, but also large string field theory 
corrections, so $g_s(\mu\a' p^+)^2\gg 1$, though $g_s^2(\mu\a' p^+)^2\ll 1$. This is a misterious regime, that has not been properly studied.

\item The largest $J$, with $J\gg \sqrt{N}/g_{YM}$, corresponds to a regime dominated by giant gravitons. Here both $\frac{J^4}{N^2}\gg 1$, but also 
$(\frac{J^4}{N^2})(\frac{g^2_{YM}N}{J^2})\gg 1$, so interacting nonplanar corrections are large. In the pp wave, $g_s \mu \a' p^+\gg 1$, and giant gravitons become 
larger than the string scale (are truly giant, therefore dominate).

\end{itemize}

We work as an expansion around the second regime, which is $\mu\a' p^+\gg 1$, but $g_s(\mu \a' p^+)^2\ll 1$ in string theory on the pp wave.

In supergravity (the $\a'\rightarrow 0$ limit of the string theory, though note that now we also have $\mu \a' p^+\gg 1$, so it is not clear that this applies 
{\em a priori}, it needs to be explicitly checked as below), the interaction term  in the action, for a closed string massless mode $\phi$, is 
\be
g_s\int d^8r \; dx^+ dx^- \phi^2\Box \phi=g_s\int d^8r \; dx^+ dx^- \phi^2(2\d_+\d_--\mu r^2\d_-^2+\d_i\d_i)\phi\;,
\ee
and gives rise to a 3-point amplitude of order
\be
{\cal A}\sim \frac{p^2\times p}{\sqrt{p_1p_2p_3}}\delta(p_3-p_1-p_2)\rightarrow \frac{1}{N}\frac{J^3}{\sqrt{J_1J_2 J_3}}\delta_{J_1+J_2,J_3}\;,
\ee
where we have written the result in SYM variables. 

\begin{figure}[ht]
\begin{center}
\epsfxsize= 5 cm \epsfbox{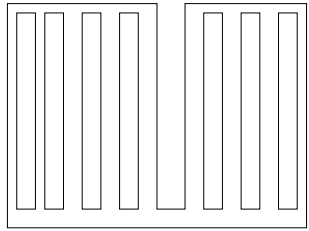}
\end{center}
\caption{Splitting and joining of 
closed strings in free SYM}\label{fig:splitdouble.eps} 
\end{figure}

This is independent on the 't Hooft coupling, so should be obtainable from a free (non-interacting) Feynman diagram calculation in SYM. Indeed that is the case:
this matches the calculation for the overlap of the free (vacuum, single-trace) operator of length $J_1$, $\Tr[Z^{J_1}]$, plus the free (vacuum, single trace)
operator of length $J_2$, $\Tr[Z^{J_2}]$, with the one of length $J=J_3=J_1+J_2$, $\Tr[Z^{J_3}]$, as in Fig.\ref{fig:splitdouble.eps}.

This diagram is indeed of order 
\be
\frac{1}{N}\frac{J^3}{\sqrt{J_1J_2 J_3}}\delta_{J_1+J_2,J_3}\;,
\ee
and moreover, the exact amplitude matches. This is the basic component of the splitting and joining of strings, i.e., of the closed string field theory, and we see that 
it is reproduced from simply a free SYM calculation, independent on the form of the interacting Hamiltonian. 

This amplitude must be understood a having the coupling $J^2/N=4\pi g_s(\mu\a' p^+)^2$, 
while the rest ($J/\sqrt{J_1J_2J_3}\delta_{J_1+J_2,J_3}$) is understood as coming from normalizations and phase space.

\begin{figure}[ht]
\begin{center}
\epsfxsize=5 cm \epsfbox{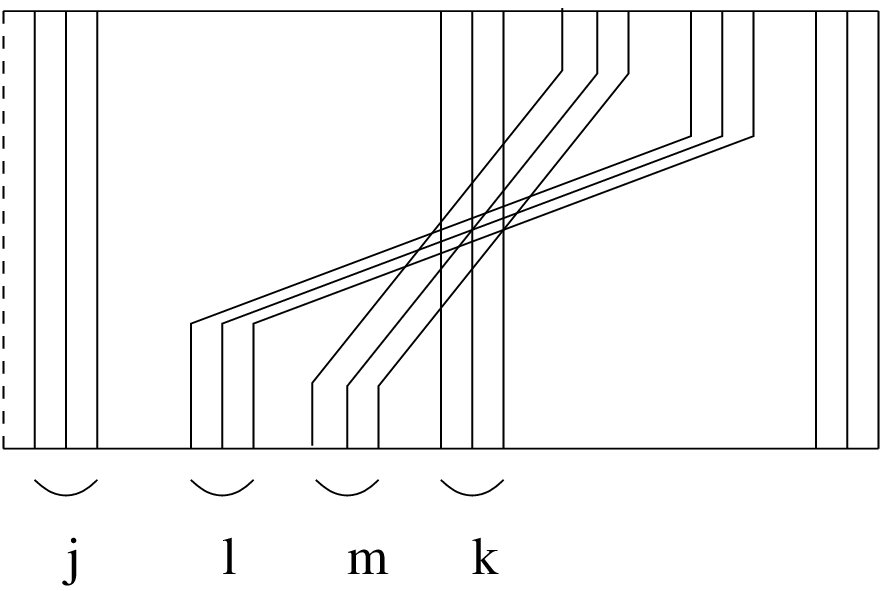}
\put(20,40){
\epsfxsize=4 cm \epsfbox{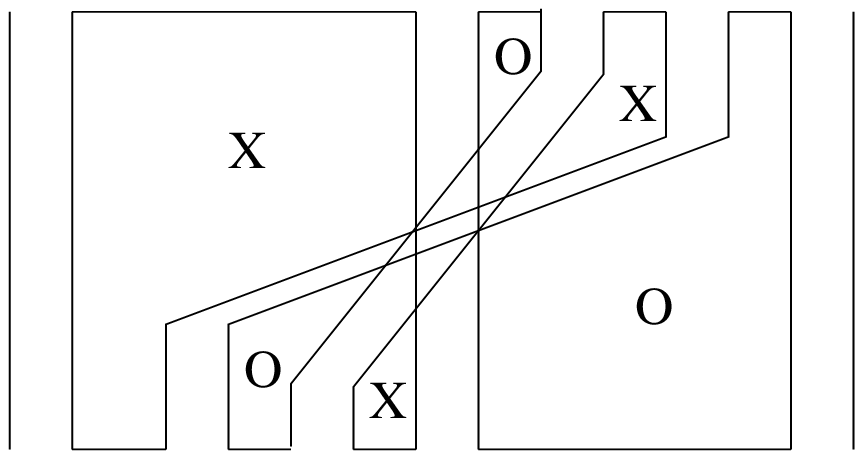} }
\end{center}
\caption{
SYM nonplanar diagram corresponding to the first string loop correction. It is a worldsheet with a handle 
made of $m+l$ bits (left figure). On the right, we see that we have a  only a factor of $1/N^2$ reduction: there is a single index loop ($N$ factor), compared with 
3 in the planar contribution.
}
\label{fig:oneloopfree}
\end{figure}

One could perhaps think that this is spurious, being restricted to supergravity computations, not the full string modes, or maybe just to free interactions (quantum mechanics
of the string field, but not string field theory). Then, to respond to this, one can consider true nonplanar interactions, that give an $J^4/N^2$ correction. The relevant diagrams
are represented in Fig.\ref{fig:oneloopfree}.

In these diagrams, for clarity we consider an insertion of an oscillator (dotted line) at the origin, to fix the cyclicity of the trace. Then the diagram hops $m$ bits 
from site $j+l$ over $k$ other bits, then $l$ bits at site $j$ over $k+m$ other bits, and are the only nonplanar diagrams at order $1/N^2$ (first nonplanarity). 
Summing over $j,l,k,m$ we obtain a factor of order $J^4$ diagrams, for a total factor of $J^4/N^2$ with respect to the free diagram. Thus indeed, we have 
diagrams of order $J^4/N^2=g_s^2(\mu\a' p^+)^4$, as advertized. 

This diagram corresponds to a term in the string theory effective action (field theory action for the string modes) of order $g_s^2(\mu\a'p^+)^4\phi^2$. 

\begin{figure}[hbt]
\centerline{
\epsfxsize 3in \epsfbox{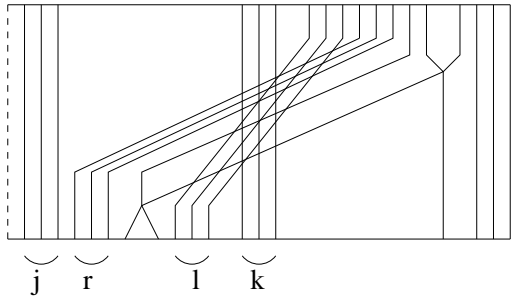}
}
\caption[]{The nonplanar generalizations of the planar rwo-vertex diagram. The handle is made up of $r+l$ bits, and the SYM interactions are on the handle.}
\label{twoloopnpl}
\end{figure}

But we might be concerned that it doesn't have the simple $g_s^2$ dependence of a string correction. It turns out that this dependence can also be obtained from 
SYM, though it is somewhat more complicated from the point of view of SYM. Indeed, we need to consider an interaction that is genuinely of order $g^4_{YM}$, so 
it contains {\em two } vertices, yet in the BMN limit, so they should be commutator interactions, of order $\frac{g^4_{YM}N^2}{J^4}$, just that with a nonplanar 
generalization, so that we get an extra $\frac{J^4}{N^2}$ factor. Such diagrams are represented in Fig.\ref{twoloopnpl}.

As we see from the diagram, the first interaction occurs at position $j+r$ (bits after the origin), and the second after another $l+k$ bits. The $r$ bits at site $j$ jump 
over $l+k$ sites, and the $l$ bits at $j+r+2$ jump over $k$ sites. This results in a handle made up of $r+l$ bits, with the two SYM interactions on the handle. 
The relevant SYM interactions are either: a) an oscillator ($\phi^m$ or $D_a$) line hopping successively over two $Z$ lines, or b) two neighbouring oscillator
lines (two $\phi$'s or two $D$'s) interacting to give two $Z$ lines, then one of the $Z$ lines, together with an neighbouring one, interact to give two other oscillator 
lines. Not surprisingly, these give the combination of the two $\frac{g^2_{YM}N}{J^2}$ factors for the commutation interactions, together with the $J^4/N^2$ of 
nonplanarity ($J^4$ for the sum over $j,r,l,k$ number of sites, the $1/N^2$ for nonplanarity), for a total of 
\be
\frac{g^4_{YM}N^2}{J^4}\frac{J^4}{N^2}=g^4_{YM}=(4\pi g_s)^2.
\ee

This then is a simple closed string loop correction on the pp wave. In \cite{Berenstein:2002sa}, a complete analysis of the interactions of SYM vertices with nonplanarity 
was done, and shown to organize in string corrections on the pp wave. To obtain the correct perturbative string field theory, it was also noticed that the correct 
contact term contributions for the string field theory action are obtained from the "string bit" approach above to the SYM diagrams, but we will not review that here.

\bibliography{AdS}
\bibliographystyle{utphys}

\end{document}